# Developing the Cyclotactor

**Staas de Jong**
LIACS
Leiden University

staas@liacs.nl

**Abstract**
This paper presents developments in the technology underlying the cyclotactor, a finger-based tactile I/O device for musical interaction. These include significant improvements both in the basic characteristics of tactile interaction and in the related (vibro)tactile sample rates, latencies, and timing precision. After presenting the new prototype's tactile output force landscape, some of the new possibilities for interaction are discussed, especially those for musical interaction with zero audio/tactile latency.

## 1. Introduction

The cyclotactor is a finger-based tactile I/O device for musical interaction, based on three main hardware components: an electromagnet, a proximity sensor, and a "keystone" component which is attached to the finger.

The keystone provides both force feedback (via a permanent magnet) and proximity input (via an infrared-reflecting surface). As opposed to a tactile display, this setup allows haptic output to directly influence haptic input. In [1], where the device was introduced, it was shown how this defining characteristic can be used to set up cyclical relationships between tactile input and output, and an example of a practical application of this to musical interaction was given.

However, a number of remaining tactile issues were identified as well, leading to the conclusion that at least the tactile output rate and analog response time of the electromagnet needed improvement. In the next Section, these and other improvements that have been made to the technology will be described. This is followed by the tactile output force landscape of the resulting prototype. Some examples of the new possibilities this leads to are then given, followed by the conclusion.

## 2. Developing the technology

### 2.1 Improvements

The hand, in resting position on a surface, was taken as the starting point for the next version of the prototype. In

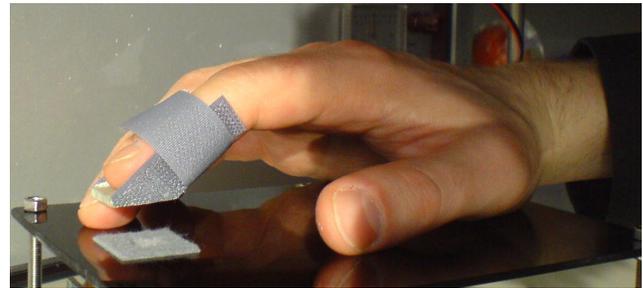
**Figure 1.** *The device surface and new keystone.*

accordance to this, the permanent magnet in the keystone was shifted to a position pressed against the fingerpad of the last finger joint (see Figure 1), instead of being attached in parallel to the general direction of the last two finger joints. This allows for a more natural tapping movement, and also resulted in a perceived improvement in the transfer of vibrotactile feedback.

In addition to this, every subcomponent of the basic system was reconsidered and improved. To begin with, the host system and hardware interface were changed to a combination of Mac OS X's CoreAudio and a Motu 828mkII, providing low-latency audio-rate voltage I/O. The output sample rate was increased from 200 Hz to 96 KHz, while improved timing precision eliminated jitter as a problem.

By adjusting the infrared (IR) detection electronics and changing to high-speed pulsed operation for the reflected IR intensity input, it was possible to double the vertical proximity range to 35 mm, while keeping sensitivity at approximately 0.2 mm. This allows for much better capturing of finger movement, at a sample rate increased from 400 Hz to 4800 Hz. The system's tactile input latency is now estimated at approximately 1.8 ms.

A number of electromagnets were built from scratch, in order to improve the response to high speed changes in coil current (and thereby magnetic output) while retaining the same output force range. In tandem with a new opamp circuit, this extended the effective output frequency range from 0-100 Hz (with a 67% amplitude drop) to 0-1000 Hz (with a more or less linear and flat response in the vibrotactually important range up to 400 Hz). The maximum added latency (to reach 99% of a magnetic target amplitude) was decreased from 33 ms to 1.0 ms, bringing the system's tactile output latency down to an estimated 2.6 ms.

New keystone prototypes based on a range of permanent magnet types were created and tested. To compare different keystone/electromagnet combinations, *maximum practicable static rejection* (*MPSR*) was used: the maximum stable rejecting force that can be encountered across the vertical distance range, without slipping sideways of the keystone becoming unavoidable.

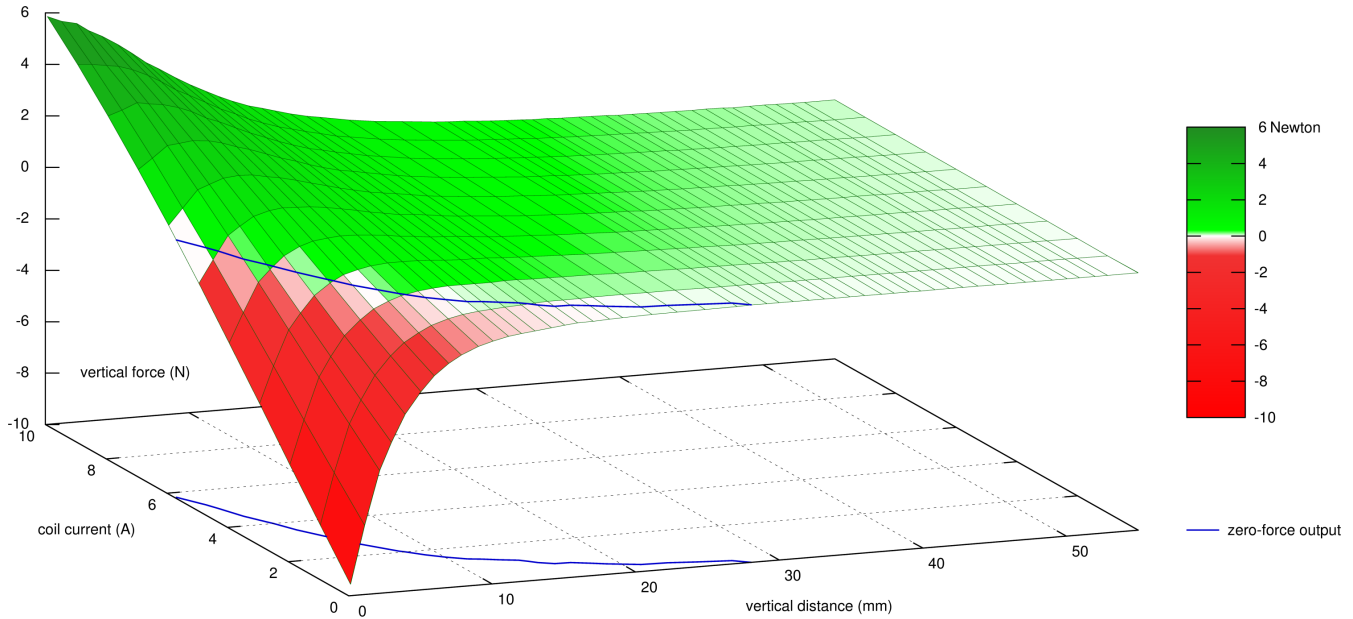

**Figure 2.** *Tactile output force landscape straight above the electromagnet's core.*

### 2.2 Tactile output force landscape

In Figure 2, the *tactile output force landscape* of the selected keystone/electromagnet combination is shown. It is directly based on measurements taken to represent the functional relationship between vertical distance, coil current, and the resulting vertical force on the keystone. This force characteristic underlies both kinesthetic and cutaneous feedback. Notably, the data set shows how fixed-current force output may well have the semantically undesirable property of first increasing, then decreasing again over distance.

## 3. Examples of new possibilities

### 3.1 Passive, touch-sensitive surface

Paradoxically, it only now becomes possible for the device to behave as an inactive surface, by actively applying the zero-force characteristic shown in Figure 2 to counteract magnetization of the electromagnet's core by the keystone. By combining this with velocity input derived from the proximity signal, a touch-sensitive surface for audio triggering can be implemented.

### 3.2 Zero audio/tactile latency

Perhaps more interestingly, the improved technology can also be used to aim for zero audio and tactile latency in percussive-type interactions. Similarly to the "preparatory gestures" mentioned in [2], a velocity input signal may again be used, to predict arrival at a certain vertical distance above the device surface. Using their known separate latencies, audio and tactile output can then each be "prescheduled" to create the equivalent of a direct, simultaneous response.

## 4. Conclusion

Significant technological improvements to the cyclotactor have been described, and a few of the resulting new possibilities for musical interaction have been hinted at. The previously discussed tactile output force landscape both enables and shows the necessity of implementing a symmetric force output signal in Newtons as the next step. This is expected to finally enable a shift in emphasis, from developing tactile technology to developing musical interactions.

## 5. Acknowledgments

I would like to thank Rene Overgauw at the Electronics Department of the Faculty of Science of Leiden University for his indispensable advice and practical support. Also thanks to Edwin van der Heide, and to the Faculty of Science's Fijnmechanische Dienst.